 \documentclass[final,5p,times,twocolumn,authoryear]{elsarticle}

\usepackage{amssymb}
\usepackage{lipsum}
\usepackage{csquotes}

\journal{Physics Letters B}

\setcitestyle{numbers,square}

\begin{document}

\begin{frontmatter}

%% Title, authors and addresses

%% use the tnoteref command within \title for footnotes;
%% use the tnotetext command for theassociated footnote;
%% use the fnref command within \author or \affiliation for footnotes;
%% use the fntext command for theassociated footnote;
%% use the corref command within \author for corresponding author footnotes;
%% use the cortext command for theassociated footnote;
%% use the ead command for the email address,
%% and the form \ead[url] for the home page:
%% \title{Title\tnoteref{label1}}

%% \tnotetext[label1]{}
%% \author{Name\corref{cor1}\fnref{label2}}
%% \ead{email address}
%% \ead[url]{home page}
%% \fntext[label2]{}
%% \cortext[cor1]{}
%% \affiliation{organization={},
%%            addressline={}, 
%%            city={},
%%            postcode={}, 
%%            state={},
%%            country={}}
%% \fntext[label3]{}

\title{Resonant suppression of the above-barrier fusion cross-section in $^{19}$O + $^{12}$C}

%% use optional labels to link authors explicitly to addresses:
%% \author[label1,label2]{}
%% \affiliation[label1]{organization={},
%%             addressline={},
%%             city={},
%%             postcode={},
%%             state={},
%%             country={}}
%%
%% \affiliation[label2]{organization={},
%%             addressline={},
%%             city={},
%%             postcode={},
%%             state={},
%%             country={}}

\author[a]{H. Desilets}
\author[a]{Rohit Kumar}
\author[a]{R.~T. deSouza}
\author[a]{S. Hudan}

\author[b]{C. Ciampi}
\author[b]{A. Chbihi}

\author[c]{K.~W. Brown}

\author[d]{K. Godbey}

\author[e]{B. Pinheiro}
\author[e]{E. N. Cardozo}
\author[e]{J. Lubian}

\affiliation[a]{%
Department of Chemistry and Center for Exploration of Energy and Matter, Indiana University
2401 Milo B. Sampson Lane, Bloomington, Indiana 47408, USA}%

\affiliation[b]{GANIL, CEA,DRF-CNRS,IN2P3, 
Blvd. Henri Becquerel, F-14076, Caen, France} %

\affiliation[c]{Facility for Rare Isotope Beams and Department of Chemistry, Michigan State University, East Lansing, MI 48823, USA} %

\affiliation[d]{
FRIB Laboratory, Michigan State University, East Lansing, Michigan 48824, USA} 

\affiliation[e]{%
Instituto de Fisica, Universidade Federal Fluminense,
Niterói 24210-340, R.J., Brazil}%

\begin{abstract}
%% Text of abstract
Fusion excitation functions for $^{19}$O and $^{19}$F + $^{12}$C were simultaneously measured along with $^{20}$O + $^{12}$C using the active-target detector MuSIC@Indiana. Examination of the cross-section just above the barrier reveals a significant suppression of fusion for the $^{19}$O + $^{12}$C system at E$_{\rm{\rm{c.m.}}}$$\sim$12 MeV. This suppression may be due to the formation of a transient $^{18}$O-n-$^{12}$C ``molecule-like'' complex favored at a particular angular momentum. Mean-field level models are used as a reference to understand the measured fusion excitation functions.
\end{abstract}

%%Graphical abstract
%\begin{graphicalabstract}
%\includegraphics{grabs}
%\end{graphicalabstract}

%%Research highlights
%\begin{highlights}
%\item Research highlight 1
%\item Research highlight 2
%\end{highlights}

\begin{keyword}
%% keywords here, in the form: keyword \sep keyword, up to a maximum of 6 keywords
Fusion \sep Radioactive beams \sep Resonant behavior in fusion \sep above-barrier fusion suppression

%% PACS codes here, in the form: \PACS code \sep code

%% MSC codes here, in the form: \MSC code \sep code
%% or \MSC[2008] code \sep code (2000 is the default)

\end{keyword}

\end{frontmatter}

%\tableofcontents

%% \linenumbers

%% main text

\section{Introduction}
\label{introduction}
The fusion of two nuclei into one is a complex process in which both the initial structure of the two nuclei and the collision dynamics can play an important role \cite{Back14,Montagnoli17,CGD15}. Nuclei are strongly interacting quantal systems that manifest both a mean field, one-body character, as well as higher-order correlations and quantal structure (shell effects). 
When two nuclei fuse, the outcome of the fusion reaction is determined by both the interaction of the two mean fields as well as the quantal nature of the two nuclei, which evolves during the collision \cite{Montagnoli17, deSouza24, Stefanini24}. 
Light nuclei in particular provide a rich testing ground for the interplay of nuclear structure and dynamics as the relative importance of nuclear structure is large. A new tool in addressing this topic is the availability of isotopic chains of radioactive nuclei at radioactive beam facilities. These beams enable a systematic examination of fusion as a function of neutron number. 
In this work, we examine the fusion of $^{18,19,20}$O incident ions with $^{12}$C nuclei. 
Specifically, the impact of two, three, or four neutrons above the doubly-magic $^{16}$O core on fusion is explored. 
\section{Experimental details}
%%\label{}
The experiment was conducted using the SPIRAL1 facility at the GANIL accelerator complex in Caen, France. 
A primary beam of $^{22}$Ne at E/A = 80 MeV bombarded a graphite target to produce a beam of $^{19}$O,$^{19}$F, and $^{19}$Ne and also a beam of $^{20}$O. These ions were accelerated by the CIME cyclotron to an energy of E/A = 2.996 MeV and then selected in B$\rho$ by the ALPHA spectrometer before being transported to the experimental setup.
The central element of the experimental setup is the active target detector MuSIC@Indiana. The beam impinged on it at an intensity up to $\sim$7$\times$10$^4$ ions/s. 

MuSIC@Indiana is a transverse-field, Frisch-gridded ionization chamber \cite{Johnstone21}. The gas in the detector serves as both a target and a detection medium. The active target approach has several intrinsic advantages over conventional thin-target measurements. It provides a direct energy and angle-integrated measure of the fusion products and allows a simultaneous measurement of multiple points on the excitation function \cite{Carnelli15}. When self-triggered, MuSIC detectors provide self-normalizing measurements since the incident beam is detected by the same detector as the reaction products. 
These advantages make use of MuSIC@Indiana a particularly effective means for measuring fusion excitation functions particularly when available beam intensities are low \cite{Johnstone21, Carnelli14}.

In this experiment, MuSIC@Indiana \cite{Johnstone21}, operated with a fill gas of CH$_4$ (99.99$\%$ purity) at pressures between 110 and 175 torr, was utilized to measure the fusion cross-section.
Ions traversing the gas volume produce a characteristic ionization that is directly measured by the segmented anode consisting of 
twenty  12.5 mm wide strips oriented transverse to the beam direction. Each strip was further subdivided left/right with respect to the beam, resulting in forty anode segments \cite{Johnstone21}. Effective collection of the ionization was ensured by operating MuSIC@Indiana at
a reduced electric field of $\sim$0.7 kV/cm/atm in the active volume \cite{Foreman81}.
 Utilizing the fast induced signal from electron motion, the cathode of MuSIC@Indiana was used to trigger the data acquisition so that an incident ion depositing at least 1.0 MeV in the detector was recorded. Charge deposited on each anode segment was processed by standard charge-sensitive and shaping amplifiers before being digitized and recorded by the data acquisition system.  
 
 To calibrate the anode energy signals, a silicon surface barrier detector was used. The measured beam energy  at the front and back of each anode strip, together with the pedestal, provided the energy calibration \cite{Johnstone21}.
 Following energy calibration, each left/right pair of anode segments was summed.
 For each incident ion the set of anode-strip energies measured is collectively referred to as a \enquote{trace}. 
 A more detailed description of the design, operation, and performance of MuSIC@Indiana can be found in \cite{Johnstone21, Johnstone22}. 
 
 \begin{figure}[h]
\includegraphics[width=0.45\textwidth]{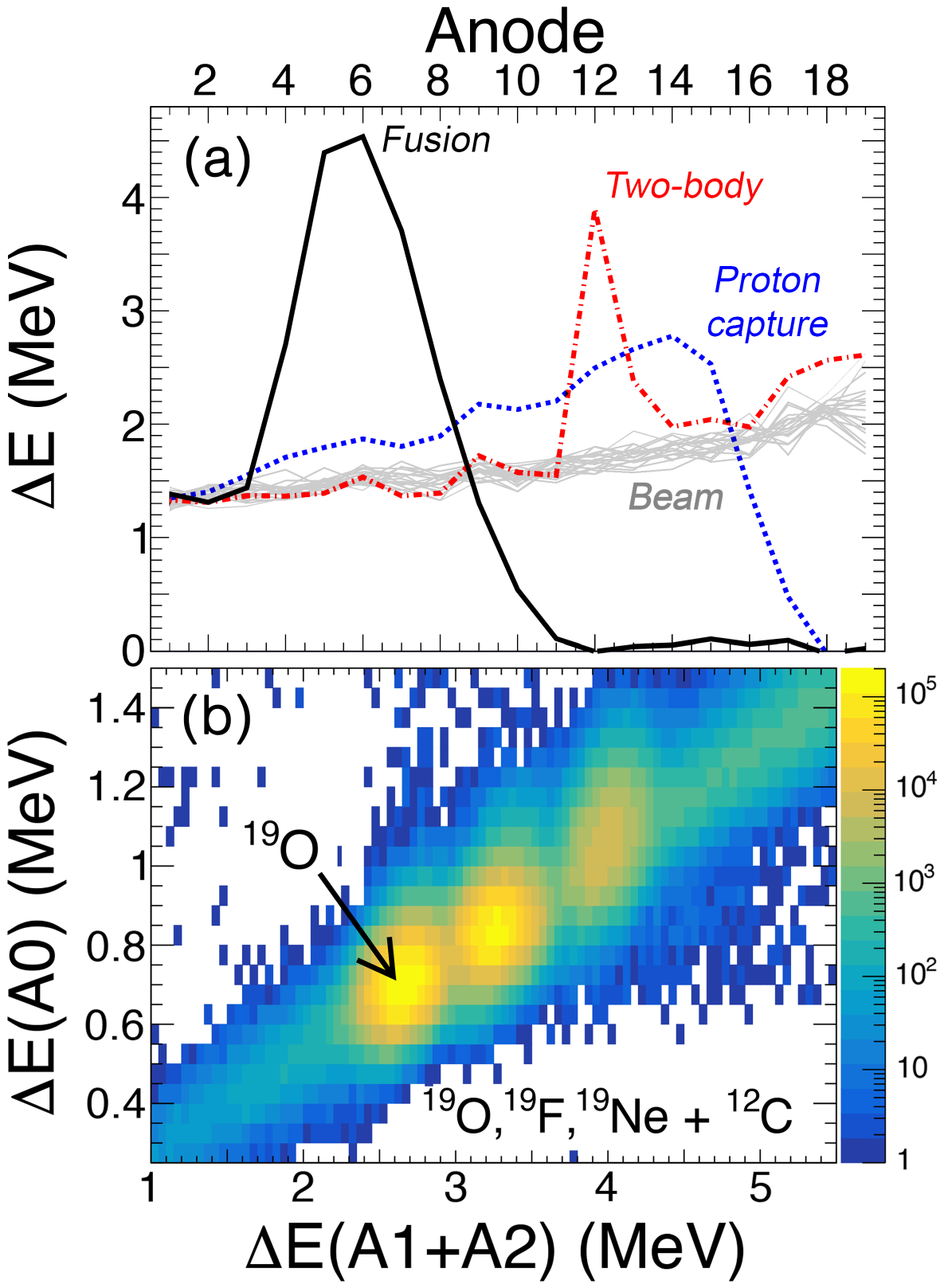}
\caption{Panel (a): Exemplary traces for fusion (black, solid), two-body (red, dash-dot), and proton capture (blue, dotted) events. Multiple beam traces are shown for reference. Panel (b): PID spectrum of the A=19 beam. Three anodes provided sufficient separation for beam selection.}
\label{fig:TraceswPIDv4}
\end{figure}

 Representative traces, indicative of three distinct processes are presented in 
Fig.~\ref{fig:TraceswPIDv4}a. The set of light-gray traces indicates the ionization due to the passage of beam ions. At the entrance of the detector an energy deposit of $\sim$1.3 MeV is observed, which increases to $\sim$2 MeV when the ion exits the detector.
In marked contrast to this ionization is the solid black trace. This trace exhibits a larger specific ionization associated with a heavy fusion product (ER: evaporation residue). For a fusion trace, a general monotonic decrease is observed in $\Delta$E after the peak. The increased ionization at anode 4 reveals the position and hence the energy at which  fusion occurred. 
Also depicted are two other reactions that occur. The dash-dot (red) trace is representative of two-body scattering events (e.g. inelastic scattering, transfer, etc.), while the dotted (blue) trace is representative of proton-capture reactions. While the two-body traces do exhibit a large ionization, they also manifest a beam-like ionization following the peak, differentiating them from fusion. Proton-capture traces exhibit increased ionization relative to the beam, however the increased ionization is modest as the change in atomic number of the traversing ion is small as compared to fusion. The characteristic traces of fusion are thus readily separated from traces associated with beam or other reaction types. Determination of the fusion cross-section involves counting the number of fusion traces relative to the number of incident ions.
Measurement of the fusion excitation function is enabled by the determination of the position at which fusion occurs \cite{Johnstone21, Carnelli14}. In the simplest analysis, the finite size of each anode strip results in an uncertainty in the energy at which the reaction occurs.

A feature of MuSIC@Indiana that distinguishes it from other MuSIC detectors is the ability to precisely insert a small silicon surface barrier detector (SBD) into the active volume from downstream. Insertion of the calibrated SBD into the active region with an accuracy of $<$0.1 mm allowed a precise measurement of the beam energy at each anode strip. The SBD provided an efficient means of performing this measurement for each pressure setting utilized. Calibration of MuSIC@Indiana with different incident ions \cite{Johnstone21} 
enabled accurate determination of the energy loss of different ions in the gas
eliminating the sensitivity to energy loss calculations, which have uncertainties as large as 15$\%$ \cite{Johnstone21, Avila17}.

The fusion cross-section, $\sigma_F$, for each anode segment is related to the number of detected ER, $N_{ER}$ by:

\begin{equation}
\sigma_F = \frac{N_{ER}}{N_{Beam}\times\Delta x}
\end{equation}

where $N_{Beam}$ is the number of incident projectile ions and $\Delta x$ is the thickness of the anode segment. All of the fusion cross-sections measured in E831 are tabulated in the Supplementary Material. The accuracy of this integrated measurement thus relies simply on the ability to distinguish ionizing events associated with fusion as compared to beam or other reaction types. Complete detection of the ER within the active volume of MuSIC@Indiana eliminates the need for an efficiency correction. Unlike thin-target measurements which are sensitive to the measured ER angular distribution,  MuSIC measurements intrinsically provide an angle-integrated measure of the cross-section. The uncertainty in the measured cross-section is largely determined by the statistical uncertainty associated with $N_{ER}$. In addition, a systematic uncertainty of 5$\%$ associated with the mis-identification of fusion events as non-fusion events for $N_{ER}$ is included.  The uncertainty associated with $N_{Beam}$ is negligible and the uncertainty associated with $\Delta x$ is defined by the pressure variation in MuSIC@Indiana which is at maximum 0.6$\%$. All these uncertainties are included in the error bars reported.

In the case of the $^{19}$O + $^{12}$C measurement, the desired $^{19}$O beam was contaminated by the presence of $^{19}$F and $^{19}$Ne.  Identification of each incident ion was accomplished by utilizing the first three anode strips of MuSIC@Indiana for a $\Delta$E$_0$-$\Delta$E$_{(1+2)}$ measurement. As evident in Fig.~\ref{fig:TraceswPIDv4}b, this measurement provided clear separation of $^{19}$O, $^{19}$F, and $^{19}$Ne ions and provided a count of each of the cleanly identified species incident on the detector. Identified ions of $^{19}$O and $^{19}$F were selected for the subsequent analysis but the statistics for $^{19}$Ne was relatively poor. In the case of the $^{20}$O beam, no contaminants were evident.

\section{Results and Discussion} 
\begin{figure}
\includegraphics[scale=0.50]{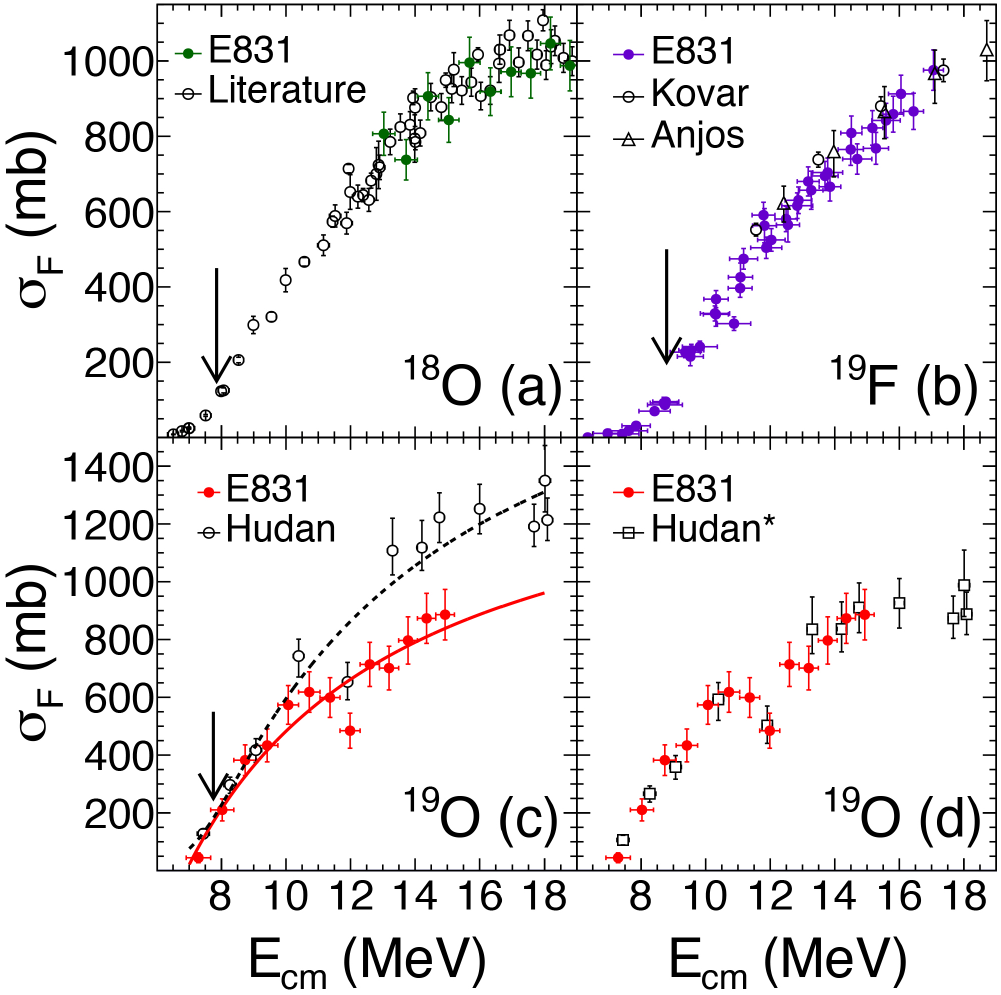}
\caption{ Comparison of the fusion excitation functions for $^{18}$O, $^{19}$O and $^{19}$F on $^{12}$C. The Bass barrier \cite{Bass74} for each system is indicated by the arrow. The previously measured cross-section for $^{19}$F + $^{12}$C is also shown for reference \cite{Kovar79, Anjos90}. Wong fits \cite{Wong73} to each of the $^{19}$O datasets are indicated by the lines.
The scaled Hudan data is depicted in panel (d) as Hudan$^*$.}
\label{fig:A19_O18}
\end{figure}

At the start of the experiment, a beam of $^{18}$O was used to confirm the operation of MuSIC@Indiana and serve as a reference measurement. The resulting excitation function is shown in Fig.~\ref{fig:A19_O18}a compared to the published excitation function from both thin- and thick-target measurements \cite{Johnstone21, Steinbach14a, Eyal76, Kovar79, Heusch82}. Good agreement of the present measurement with the literature is observed.
Presented in Fig.~\ref{fig:A19_O18}b,c are the fusion excitation functions for 
$^{19}$F and $^{19}$O ions on $^{12}$C. For reference, the systematic barriers calculated with the Bass model \cite{Bass74} are indicated by the arrows. In the case of $^{19}$F, the present data are compared with the published thin-target results of Kovar et al. \cite{Kovar79} and Anjos et al. \cite{Anjos90}. While the agreement is quite good, the present measurement is systematically $\sim$5-10$\%$ lower than those reported by Kovar et al. 
As the present measurement is self-normalizing and 
Kovar reports a systematic uncertainty of 5$\%$ \cite{Kovar79}, we regard the present measurement to be more accurate. 

The striking feature of Fig.~\ref{fig:A19_O18} is the significant reduction of the fusion cross-section, $\sigma_F$, for $^{19}$O at E$_{\rm{\rm{c.m.}}}$$\sim$12 MeV, presented in Fig.~\ref{fig:A19_O18}c,d. The difference in $\sigma_F$ between the low point and adjacent points exceeds the uncertainty of the measurement. If compared to a smooth interpolation between points at E$_{\rm{\rm{c.m.}}}$$\sim$11.4 MeV and 12.6 MeV, the cross-section is suppressed at E$_{\rm{\rm{c.m.}}}$=12 MeV by $\sim$180 mb. For the
$^{19}$F data a single point at E$_{\rm{c.m.}}$=10.9 MeV manifests an unexpectedly low cross-section.

To verify the accuracy of this suppression for $^{19}$O, we compare the measured cross-section to previously measured data~\cite{Hudan20} in Fig.~\ref{fig:A19_O18}c. 
This previously measured data \cite{Hudan20, Singh17} was a thin-target measurement that utilized an Energy/Time-of-Flight (ETOF) approach to identify evaporation residues and extract the fusion cross-section. 
The significantly smaller horizontal error bars in the Hudan analysis reflect the thin-target nature of the measurement. For E$_{\rm{c.m.}}$$<$ 10 MeV, the two datasets are in good agreement, while for E$_{\rm{c.m.}}$$>$ 13 MeV, the results from Hudan et al. exceed the present measurement by as much as $\sim$34$\%$.
Despite the difference between the two datasets for E$_{\rm{c.m.}}$$>$ 10 MeV,
the two datasets, which utilize different techniques to measure the fusion cross-section, independently manifest a suppression of $\sigma_F$ at E$_{\rm{c.m.}}$$\sim$12 MeV. This agreement confirms the existence of a suppression of the fusion cross-section at E$_{\rm{c.m.}}$$\sim$12 MeV.

Extraction of the fusion cross-section in ETOF approach requires knowledge of the efficiency of the experimental setup. Incomplete angular coverage in the ETOF measurement necessitates use of a statistical model to calculate the efficiency, which is decay channel dependent. 
A change in the competition between different decay channels as a function of excitation energy alters the detection efficiency. 
Accurate prediction of the competition between proton, neutron, and $\alpha$-particle emission is necessary to correctly calculate the detection efficiency. Significant underestimation of $\alpha$ emission by the statistical model in fusion of $^{18}$O + $^{12}$C has been previously observed \cite{Vadas15}.
The agreement between the two datasets at the three lowest energies suggests that in this energy regime, the efficiency is correctly predicted. 
The over-prediction by Hudan et al. at higher energies might signal an incorrect estimation of the efficiency possibly due, at least in partialy, to an incorrect prediction of the decay sequence. 

Alternatively, the increased cross-section above 10 MeV might indicate the opening of a new reaction channel for $^{19}$O, one that the ETOF measurement is sensitive to and identifies as fusion. The different sensitivities of the ETOF and MuSIC techniques in measuring fusion should be appreciated. A thorough re-examination of the uncertainties associated with extracting the fusion cross-section from ETOF data for $^{19}$O is presently underway. Independent of this discrepancy in the cross-section above 10 MeV, it is crucial to appreciate that both datasets independently manifest a suppression of $\sigma_F$ at $\sim$12 MeV.

To jointly utilize the data from both $^{19}$O datasets, each dataset was independently fit with a function that describes the penetrability of an inverted parabola (Wong formula) \cite{Wong73}. The resulting fits, displayed as the lines in Fig.~\ref{fig:A19_O18}c, were then used to scale the Hudan data to the present experiment. The combined data is presented in Fig.~\ref{fig:A19_O18}d. As the fits simply serve the purpose of re-scaling the Hudan data to the present dataset, no physical interpretation is ascribed to the scaling factor. 

\begin{figure}[h]
\begin{center}
\includegraphics[scale=0.45]{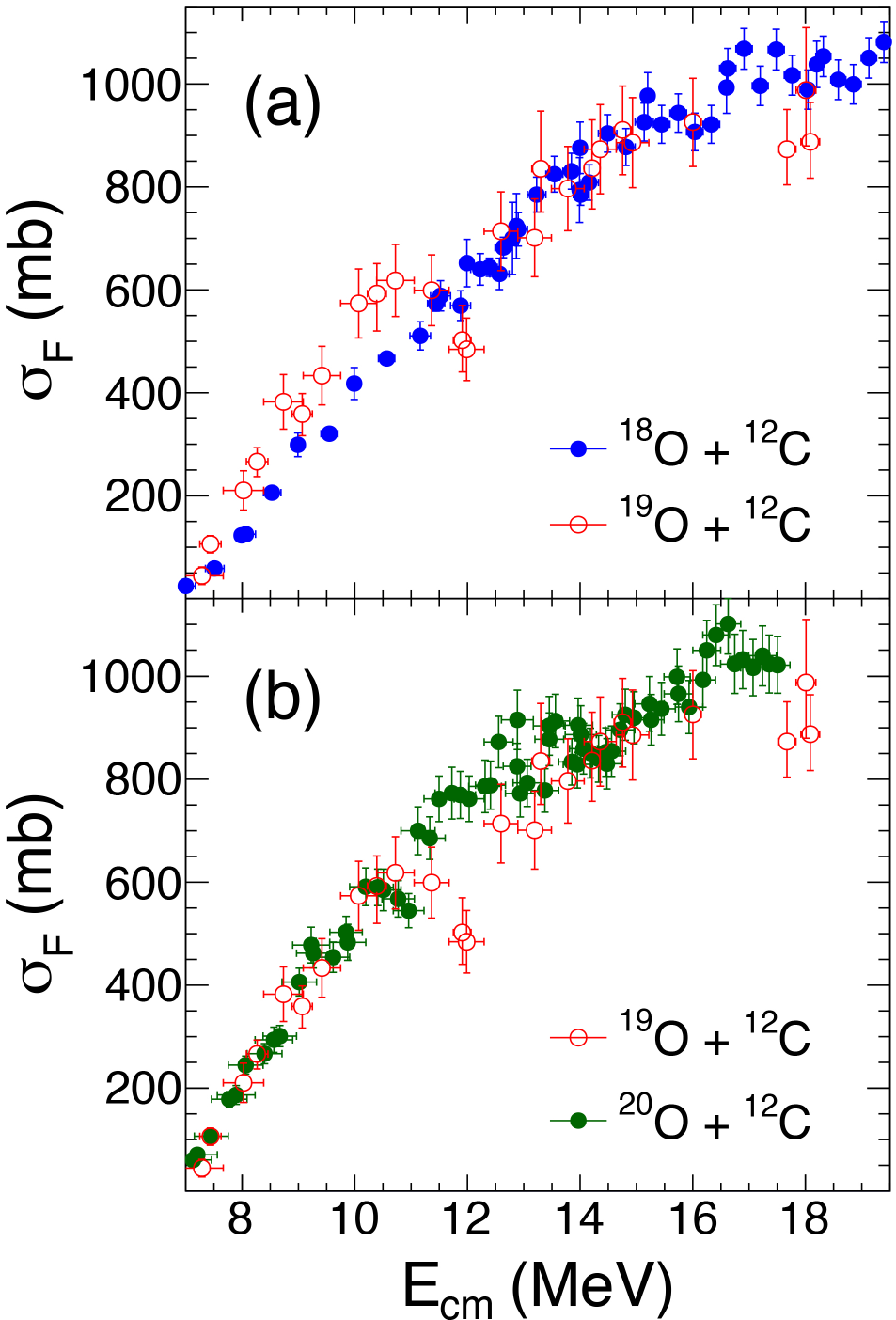}
\caption{
Comparison of the fusion excitation functions for $^{18}$O,$^{19}$O, and $^{20}$O + $^{12}$C.
}
\label{fig:18_19_20_EF}
\end{center}
\end{figure}

Comparison of the fusion excitation functions for $^{18,19,20}$O + $^{12}$C in Fig.~\ref{fig:18_19_20_EF} reveals a remarkable behavior. For low energies, E$_{\rm{c.m.}}$$<$ 11 MeV, one observes in Fig.~\ref{fig:18_19_20_EF}a that as compared to the $^{18}$O data, the $^{19}$O data manifests a slightly larger cross-section for a given energy. 
This behavior can be understood as the impact of a slightly reduced barrier (larger fusion radius) due to the presence of the additional unpaired neutron that is more loosely bound. 
At higher E$_{\rm{c.m.}}$$>$ 13 MeV energies, the cross-section for the two incident isotopes is essentially the same. Between these two regimes for E$_{\rm{c.m.}}$$\sim$ 12 MeV, one observes the dramatic suppression of the fusion cross-section in $^{19}$O.

The fusion cross-section in $^{20}$O + $^{12}$C reported in \cite{Hudan24} was extended down to E$_{\rm{c.m.}}$$\sim$ 6 MeV in this analysis.
In Fig.~\ref{fig:18_19_20_EF}b, one observes that the excitation function for $^{20}$O is very similar to that of $^{19}$O at both low and high energies. 
Close examination of the $^{20}$O excitation function from 10 MeV$<$E$_{\rm{c.m.}}$$<$12 MeV is suggestive of a slight suppression of fusion for E$_{\rm{c.m.}}$$\sim$ 11 MeV. 
While the magnitude of this suppression for $^{20}$O would by itself perhaps not warrant attention, the existence of the strong suppression in $^{19}$O might indicate that a similar resonance exists in $^{20}$O, but is quenched.

\begin{figure}[h]
\begin{center}
\includegraphics[scale=0.50]{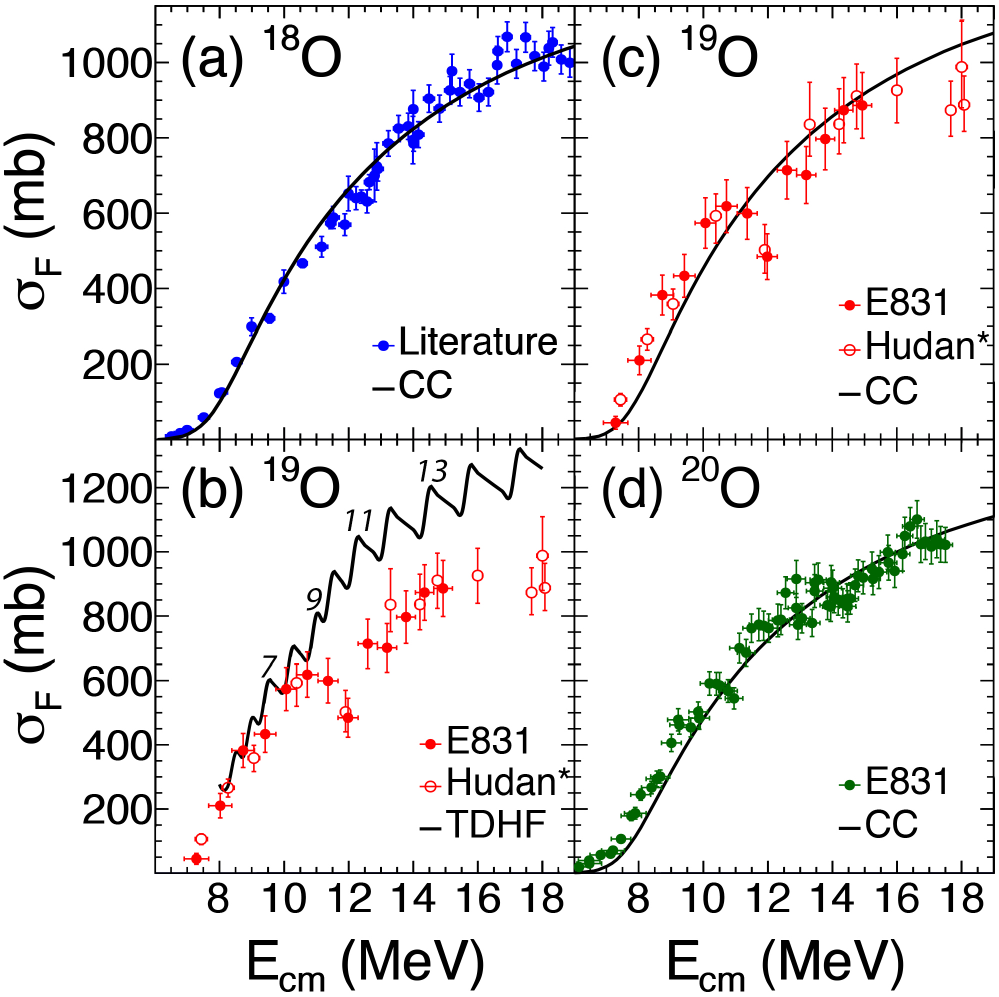}
\caption{
Comparison of the experimental fusion excitation functions with results from TDHF and CC models.
}
\label{fig:comp_models}
\end{center}
\end{figure}

To better understand the behavior of these excitation functions, we compare the experimental data with both time-dependent Hartree-Fock (TDHF) and coupled channels (CC) microscopic calculations. Shown in Fig.~\ref{fig:comp_models} are the experimental data juxtaposed with the theoretical predictions. Both models allow us to explore the mean-field behavior of the fusion cross-section for the systems considered. On general grounds,  TDHF is a microscopic, many-body approach that is well suited to describe the large-amplitude collective motion associated with fusion processes while also describing the dynamics of transfer and equilibration between the collision partners~\cite{Godbey17,Simenel20}.
Effects from Pauli blocking are also naturally included, resulting in a more realistic picture of the fusion dynamics in fermionic systems~\cite{Simenel17,Umar21}.
These effects are included self-consistently and no fits are performed beyond the original calibration of the energy density functional (EDF), representing the inter-nucleon interaction, to static properties of nuclei. The TDHF approach then provides a parameter-free means to explore the reaction dynamics, with the only input besides the interaction being the collision geometry and initial energy of the systems.
Fusion calculations are performed for different $\ell$-waves, resulting in a saw-tooth behavior of the fusion cross-section with energy \cite{deSouza24, Esbensen12,Simenel13}. This saw-tooth, behavior in contrast to previous TDHF calculations \cite{Hudan20}, properly accounts for the quantization of angular momentum between the colliding nuclei.
In Fig.~\ref{fig:comp_models}b for E$_{\rm{c.m.}}$$<$10 MeV, the TDHF calculations are in good agreement with the experimental cross-sections. As was seen in the fusion of lighter oxygen isotopes, with increasing incident energy the TDHF calculations overpredict the measured data \cite{deSouza24}. The suppression of the fusion cross-section observed in the experimental data corresponds to the barrier associated with $\ell$$\sim$10$\hbar$.

Also presented in Fig.~\ref{fig:comp_models} are the results of CC calculations using the FRESCO code~\cite{THOMPSON1988167}. Coupled channel approaches have been quite successful in describing the light-ion reactions~\cite{Esbensen12,Asher21a}. The ground state density distributions of the nuclei in this model are taken from Dirac-Hartree-Bogoliubov (DHB) calculations. As such, they best represent the ground-state density distributions at the mean-field level. The real part of the optical potential is derived by double-folding the projectile and target densities. Using a zero-range approximation for the interaction, the São Paulo potential is obtained~\cite{CPH97,CCG02}. For the imaginary part, a short-range Woods-Saxon potential is used to absorb the flux tunneling or overcoming the Coulomb barrier. The potential parameters are $W = 50$ MeV, $r_W =$ 1.06 fm, and $a_W = $ 0.2 fm. In addition to the ground state, the impact of the first excited state of both colliding nuclei is included in the calculation and indicated by the solid line. Although the high-lying first excited state of $^{12}$C has a negligible effect, including the first excited state for the oxygen does result in a small increase of the fusion cross-section. The role of higher-lying states can be disregarded as they have a diminishing impact on the fusion cross-section. In the case of $^{18}$O, presented in Fig.~\ref{fig:comp_models}a, the CC calculations provide a generally good description of the experimental data over the entire energy range shown (around and above the barrier). While small oscillations, likely due to different $\ell$-wave barriers, are not reproduced, the average fusion cross-section from the CC model is in remarkably good agreement with the experimental data.

In contrast to the results for $^{18}$O, the comparison of CC calculations with the experimental data for $^{19}$O and $^{20}$O is quite interesting. At low energies, E$_{\rm{c.m.}}$$<$10 MeV, the CC fusion cross-section is slightly lower than the experimental data. On the other hand, for E$_{\rm{c.m.}}$$>$13 MeV, the CC calculations provide a reasonably good description of the experimental cross-sections. Most notably, the significant fusion suppression observed for $^{19}$O at E$_{\rm{c.m.}}$$\sim$12 MeV is not predicted. 

On general grounds, the dramatic suppression of fusion at a particular energy for $^{19}$O could be understood as a threshold for opening a new reaction channel - one that provides an alternate path for the reaction flux as compared to fusion. For $^{19}$O this reaction channel opens at E$_{\rm{c.m.}}$$\sim$12 MeV. We hypothesize that resonant formation of a molecule-like complex consisting of $^{18}$O-n-$^{12}$C which subsequently breaks up without fusing could play a role. Asher et al. previously hypothesized a similar \enquote{molecule-like} resonance in the fusion of $^{17}$O + $^{12}$C at  E$_{\rm{c.m.}}$$\sim$14 MeV \cite{Asher21a}. However, a subsequent high-resolution measurement of the fusion excitation function for $^{17}$O reduced the magnitude of the observed suppression \cite{Hudan23} to a level approximately consistent with the presence of different $\ell$-wave barriers \cite{Esbensen12, Simenel13}, weakening the argument for such a molecular structure. The dip in $^{19}$O differs in both magnitude ($\sim$180 mb) and position (E$_{\rm{c.m.}}$$\sim$12 MeV) as compared to $^{17}$O ($\sim$50 mb;$\sim$14 MeV). 

Formation of this \enquote{molecule-like} resonant state, which does not result in fusion, could be sensitive to the incoming $\ell$-wave of the colliding nuclei. The low neutron separation energy of 3956 keV for $^{19}$O relative to $^{18}$O and $^{20}$O, 8045 keV and 7608 keV respectively, might aid formation of the complex. The absence of a suppression of the cross-section for $^{18}$O, along with a very small suppression signature for $^{20}$O, could indicate that neutron pairing in the {\em sd} shell hinders the formation of the resonance and the subsequent suppression of the cross-section.
In comparing systems with unpaired neutrons, one might expect an increased suppression and decrease in the energy at which it occurs as the number of valence neutrons increases, motivating future measurement of fusion for $^{21}$O+$^{12}$C to better understand the systematics along the isotopic chain.
Determining whether this hypothesis for forming a $^{18}$O-n-$^{12}$C complex is valid will require a dedicated experiment that measures not only fusion but other competing reaction channels.

\section{Conclusions}

The fusion excitation functions for $^{19,20}$O + $^{12}$C were measured with an active-target technique. This self-normalizing, energy- and angle-integrated measurement allowed a detailed comparison of these high-resolution fusion excitation functions with that of $^{18}$O + $^{12}$C. 
A significant depletion of the fusion cross-section at E$_{\rm{c.m.}}$$\sim$12 MeV was observed in $^{19}$O + $^{12}$C. 
 An independent thin-target measurement utilizing the ETOF technique to identify fusion products confirms the suppression of fusion at this energy. The suppression observed for $^{19}$O was not observed for the fusion of $^{18}$O, and only a very slight suppression was observed for $^{20}$O.

Though the TDHF model, with the inclusion of $\ell$-waves, provides a good description of the $^{19}$O excitation function for E$_{\rm{c.m.}}$$<$10 MeV, for larger energies it systematically over-predicts the experimental data. While the CC calculations provide a generally good description of the excitation function over the entire energy range measured, for the neutron-rich $^{19}$O and $^{20}$O nuclei, a slight under-prediction is observed at lower energies, E$_{\rm{c.m.}}$$<$11 MeV. Neither mean-field model reproduces the significant fusion suppression observed for $^{19}$O.
 
It is hypothesized that this suppression of the near-barrier fusion cross-section in $^{19}$O could indicate the formation of a \enquote{molecule-like} complex of $^{18}$O-n-$^{12}$C at a particular energy and angular momentum that does not result in fusion. Formation of this transient complex could depend sensitively on the number of valence neutrons in the {\em sd} shell and their pairing, though more detailed simulations and more comprehensive experiments are required to unambiguously identify the source and evolution of the suppression.

\section*{Acknowledgements}
We acknowledge the high-quality beam and experimental support provided by the technical and scientific staff at the Grand Acc\'{e}l\'{e}rateur National d'Ions Lourds (GANIL), in particular D. Allal, B. Jacquot, and D. Gruyer.
We are thankful for the high-quality services of the Mechanical Instrument Services and Electronic Instrument Services facilities at Indiana University.
We would also like to thank Witold Nazarewicz for providing feedback and discussions on the present work.
This work was supported by the U.S. Department of Energy Office of Science under Grant No. 
DE-SC0025230 and Indiana University. The research leading to these results has received funding from the European Union's HORIZON EUROPE Program under grant agreement no. 101057511. R. deSouza  gratefully acknowledges the support of the GANIL Visiting Scientist Program. Kyle Brown acknowledges support from Michigan State University. K. Godbey acknowledges support by the Department of Energy under Award Numbers DOE-DENA0004074 (NNSA, the Stewardship Science Academic Alliances program) and DE-SC0023175 (Office of Science, NUCLEI SciDAC-5 collaboration). Brazilian authors were supported in part by local funding agencies CNPq, FAPERJ, CAPES, and INCT-FNA (Instituto Nacional de Ci\^{e}ncia e Tecnologia, F\'{i}sica Nuclear e Aplica\c{c}\~{o}es), research Project No. 464898/2014-5.

\textit{Data Availability Statements}. 
The supporting data for this article are from the e831\_21 experiment and are registered as https://doi.org:10.10.26143/ganil-2023-e831\_21 following the GANIL Data Policy.

%% The Appendices part is started with the command \appendix;
%% appendix sections are then done as normal sections

%% If you have bibdatabase file and want bibtex to generate the
%% bibitems, please use
%%

%\bibliographystyle{elsarticle-harv} 

\bibliographystyle{elsarticle-num} 
%\bibliography{O19xsect_v3}

%% else use the following coding to input the bibitems directly in the
%% TeX file.

%%\begin{thebibliography}{00}

%% \bibitem[Author(year)]{label}
%% For example:

%% \bibitem[Aladro et al.(2015)]{Aladro15} Aladro, R., Martín, S., Riquelme, D., et al. 2015, \aas, 579, A101

%%\end{thebibliography}

\end{document}